\documentclass[graybox]{svmult}

\usepackage{mathptmx}       
\usepackage{helvet}         
\usepackage{courier}        
\usepackage{type1cm}        
\usepackage{makeidx}         
\usepackage{graphicx}        
\usepackage{multicol}        
\usepackage[bottom]{footmisc}

\makeindex             

\def\lsim{\mathrel{\mathop
  {\hbox{\lower0.5ex\hbox{$\sim$}\kern-0.8em\lower-0.7ex\hbox{$<$}}}}}
\def\gsim{\mathrel{\mathop
  {\hbox{\lower0.5ex\hbox{$\sim$}\kern-0.8em\lower-0.7ex\hbox{$>$}}}}}

\def\apj{ApJ}

\def\apjs{ApJS}

\def\PhysRev{PhysRev}

\def\mnras{MNRAS}

\def\prd{PRD}

\def\xisp{\xi(\sigma, \pi)}

\def\xips{\xi(\pi,\sigma)}

\newcommand{\nc}{\newcommand}

\nc{\be}[1]{\begin{equation}\mbox{$\label{#1}$}}
\nc{\bea}[1]{\begin{eqnarray}\mbox{$\label{#1}$}}

\nc{\Section}[2]{\section{#2}\label{#1}}
\nc{\Bibitem}[1]{\bibitem{#1}}
\nc{\Label}[1]{\label{#1}}

\nc{\Mpc}{Mpc/h}
\nc{\vev}[1]{\langle #1 \rangle}

\nc{\calH}{{\cal{H}}}

\nc{\eea}{\end{eqnarray}}
\nc{\ee}{\end{equation}}

\begin{document}

\title*{The anisotropic redshift space galaxy correlation function:  detection on the BAO Ring }
\author{Enrique Gaztanaga \& Anna Cabre}
\institute{Enrique Gaztanaga \at Institut de Ciencies de l'Espai (IEEC/CSIC), www.ice.cat,
Barcelona  \email{gazta@ice.cat}
\and
Anna Cabre \at Institut de Ciencies de l'Espai (IEEC/CSIC), www.ice.cat,
Barcelona  \email{cabre@ice.cat}}
%
%
\maketitle

\abstract*{
In a series of papers we have recently studied 
the clustering of LRG galaxies in the latest spectroscopic SDSS 
data release,  which has 75000 LRG galaxies sampling 1.1 ${\,\rm Gpc^3/h^3}$
to z=0.47. Here we focus on detecting a local maxima shaped as a
circular ring  in the bidimensional galaxy correlation function $\xisp$, 
separated in perpendicular $\sigma$ and line-of-sight $\pi$ distances.
We find a significant detection of such a peak at $r\simeq 110$Mpc/h.
The overall shape and location of the ring is consistent with
it originating from the recombination-epoch baryon acoustic oscillations (BAO). This
agreement provides support for the current understanding of how large scale structure
forms in the universe.  We study the significance of such feature using large mock 
galaxy simulations  to provide  accurate errorbars.}

\abstract{
In a series of papers we have recently studied 
the clustering of LRG galaxies in the latest spectroscopic SDSS 
data release,  which has 75000 LRG galaxies sampling 1.1 ${\,\rm Gpc^3/h^3}$
to z=0.47. Here we focus on detecting a local maxima shaped as a
circular ring  in the bidimensional galaxy correlation function $\xisp$, 
separated in perpendicular $\sigma$ and line-of-sight $\pi$ distances.
We find a significant detection of such a peak at $r\simeq 110$Mpc/h.
The overall shape and location of the ring is consistent with
it originating from the recombination-epoch baryon acoustic oscillations (BAO). This
agreement provides support for the current understanding of how large scale structure
forms in the universe.  We study the significance of such feature using large mock 
galaxy simulations  to provide  accurate errorbars.}

\section{Gravitational instability}

Is the large scale structure that we see in the galaxy distribution 
produced by gravitational growth from some small initial  fluctuations?
 We will explore two ways of addressing this
question with measurements of  the 2-point galaxy correlation:
\be{x2}
\xi(r,t) = \langle \delta(r_1,t) \delta(r_2,t) \rangle 
\ee
where $r=|r_2-r_1|$ and $\delta(r) = \rho(r)/\bar{\rho}-1$ is the local density fluctuation about 
the mean $\bar{\rho}=\langle\rho\rangle$, and the expectation values are taken over different
realizations of the model or physical process. In practice, the expectation
value is over different spatial regions in our Universe, which are
assumed to be a fair sample of possible realizations (see Peebles 1980).
The measured redshift distance of a galaxy differs from the true 
radial distance  by its peculiar velocity along the line-of-sight.   
We can split the distance $\vec{r}$ into its component along the
line-of-sight (LOS) $\pi$ and perpendicular to the LOS $\sigma$,
where $r^2 = \pi^2 + \sigma^2$. Azimuthal symmetry 
implies $\xi$ is in general a function of $\pi$ and $\sigma$ alone: $\xisp$.
 
Consider the fully non-linear fluid equations  that determine the gravitational evolution of
density fluctuations, $\delta$,  and the divergence of the
velocity field, $\theta$, in an expanding universe for a pressureless irrotational fluid. 
In Fourier space (see Eq.37-38 in \cite{bernar}):
\begin{eqnarray}
&\dot{\delta}+\theta = -\int dk_1 dk_2 ~\alpha(k_1,k_2)\theta(k_1)\delta(k_2)
 \label{eq:fluid}&
\\
&\dot{\theta}+\calH\theta+{3\over{2}} \Omega_m H^2\delta 
= -\int dk_1 dk_2 ~\beta(k_1,k_2)\theta(k_1)\theta(k_2) &
\nonumber
\end{eqnarray}
where derivatives are over conformal time $d\tau\equiv dt/a$ and $\calH(\tau) \equiv d\ln a/d\tau= a H$
is given by the expansion rate $H=\dot{a}/a$ of the cosmological scale factor $a$.
On the left hand side $\delta=\delta(k,\tau)$ and $\theta=\theta(k,\tau)$
are functions of the Fourier wave vector $k$.
The integrals are over vectors $k_1$ and $k_2$ constrained to $k= k_2-k_1$.
The right hand side of the equation include the non-linear terms which are
quadratic in the field and contain the mode coupling functions $\alpha$ and $\beta$.
When fluctuations are small we can neglect  the quadratic terms in the equations and
we then obtain the linear solution $\delta_L$. The first
equation yields $\dot{\delta_L}=-\theta_L$, which combined 
with the second equation yields the well known harmonic oscillator
equation for the linear growth:
\begin{equation}
\ddot{\delta}_L+\calH\dot{\delta}_L-{3\over{2}} \Omega_m \calH^2\delta_L = 0
\label{eq:harmonic}
\end{equation}
Because the Fourier transformation is linear,  this  equation is valid in Fourier or in configuration space.
In linear theory
each Fourier mode and each local fluctuation evolves independently of the others, moreover,
they all grow linearly out of the initial fields with the same growth function,
ie $\delta_L(t) = D(t) \delta_0$, where $\delta_0$ is the value of the
field at  a point (or a given Fourier mode) at some initial time 
and $D(t)$ is the linear growth function, which is a solution
to the above harmonic equation. In a flat universe dominated by
cold dark matter (CDM) we have that $H \propto a^{-3/2}$ and
$D(t)$ goes as the scale factor $D(t) \propto a$. In an accelerated phase,
such as $\Lambda$CDM, the growth halts or grows less rapidly with $a$.
Thus measurements of $D(t)$ can  be used as an independent diagnostics for accelerated expansion.

\subsection{BAO signature}

\begin{figure}[t]
\sidecaption[t]
\includegraphics[scale=.3]{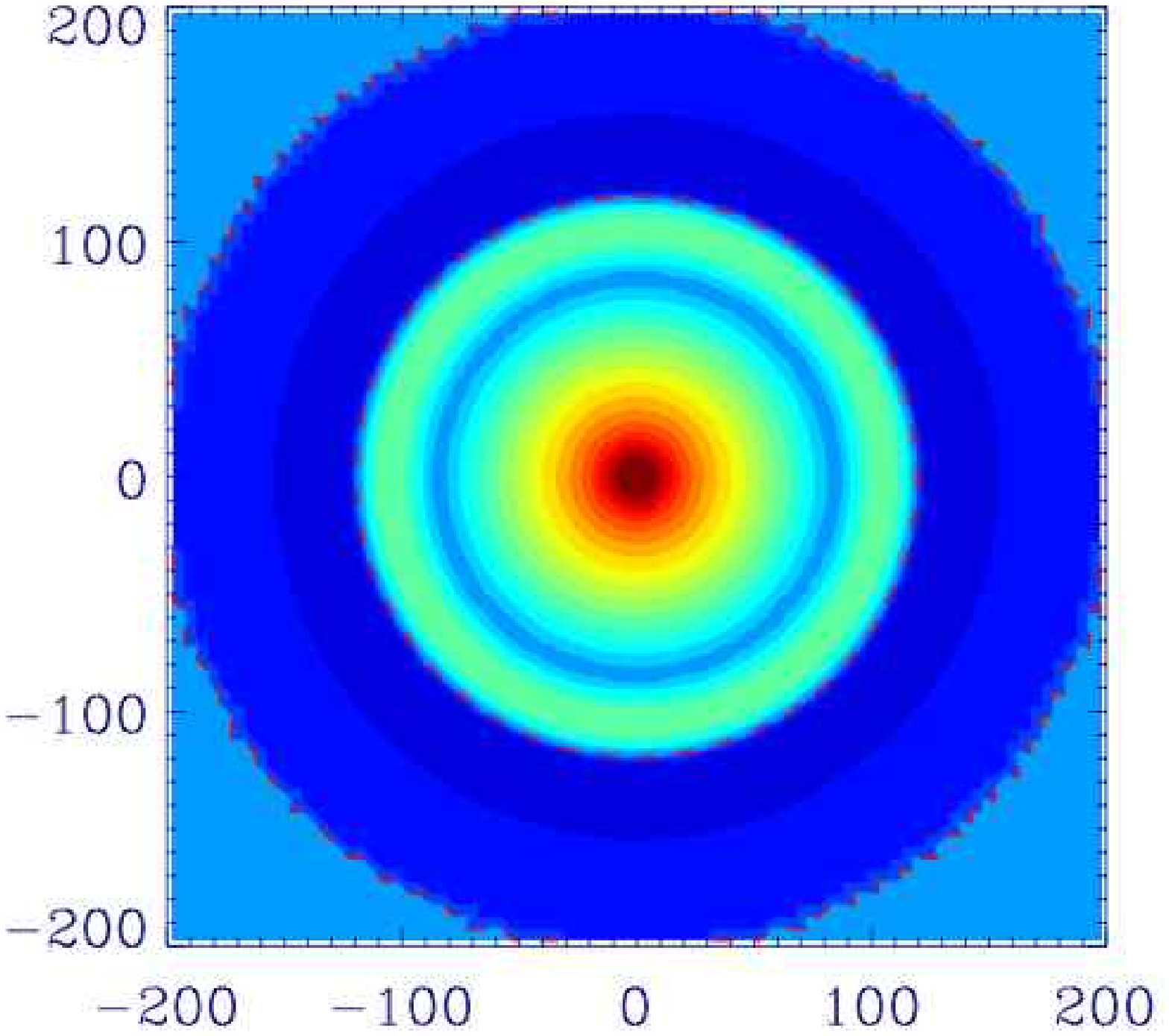}
\includegraphics[scale=.6]{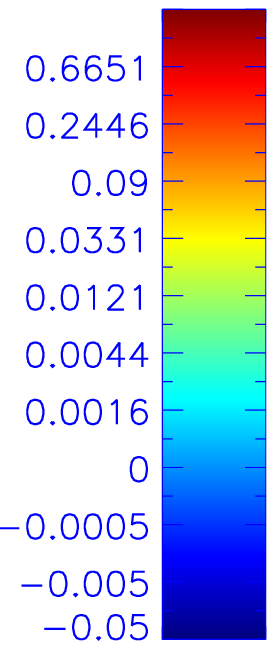}
\caption{The prediction of $\xisp$ without redshift space distortions.
The vertical axis shows the radial direction, $\pi$, while the horizontal panel shows
the transverse direction $\sigma$. There is  a prominent 
local maximum corresponding to the  BAO ring 
at a radius of about $110 \Mpc$ and $\simeq 20 \Mpc$ thickness
 (wide green circle between blue rings).
\label{fig:pisigma0}}
\end{figure}

Consider now our observable, 
the 2-point function in Eq.[\ref{x2}]. In linear theory:
 $\xi(r,t) = D(t)^2 \xi(r,0)$. This means that on large scales, ie $r>10 \Mpc$, where
 $\delta<1 $, we then expect the shape of $\xi(r)$ today to be the same as the shape
 it took in the early universe. The prediction, ignoring redshift space distortions,
 is shown in  Fig.\ref{fig:pisigma0}. The BAO ring corresponds to the
 local maxima at a radius of about $110 \Mpc$.
 This will be tested here by comparing the linear theory
 prediction for the baryon acoustic oscillation (BAO) with the measurements in
 the large scale galaxy distribution.  

The mean BAO signature in the 2-point correlation has been detected in Luminous Red Galaxies (LRG's) of 
the SDSS galaxy  survey   \cite{detection} using
the monopole, ie the average signal of $\xips$ in circles of constant $\pi^2+\sigma^2$.
In \cite{paper1,paper4} and below  we will show that the galaxy distribution 
 has a BAO ring  very similar to that predicted in the initial conditions 
 of the  CDM model.

\subsection{Growth factor}
 
 On the other hand, if we knew the shape and amplitude of the initial 
 conditions $\xi(r,0)$, we could then
estimate $D(t)^2 \simeq \xi(r,t)/\xi(r,0)$ from measurements of $\xi(r,t)$ and
compare it to the linear solution of Eq.[\ref{eq:harmonic}] to test
gravitational instability. But this approach is difficult in practice because there is a bias
in the amplitude of galaxy clustering compare to the one in dark matter fluctuations.
We could instead  test gravitational growth with independence of time
or initial conditions by using the linear relation between density and
velocities,  $\dot{\delta}_L=-\theta_L =   f\delta_L$, where $f \equiv \dot{D}/D = d\ln{D}/d\ln{a}$
is the velocity growth factor. For a flat dark matter dominated universe $D=a$ and $f=1$,
while for a flat accelerated universe $f= \Omega_m(z)^\gamma <1$, 
where $\gamma$ is the
gravitational growth index and $\Omega_m(z)$ is the matter density
at a redshift z where $a=1/(1+z)$.
The growth index  separates out two
physical effects on the growth of structure: $\Omega_m(z)$ depends on the expansion
history while $\gamma$ depends on the underlaying theory of gravity  \cite{linder1}. 
The value $\gamma=0.55$ corresponds to standard gravity, while $\gamma$ is different for modified
gravity, for example $\gamma=0.68$ in the braneworld cosmology. We will show next
how $f$ can be measured using redshift space distortions in the galaxy correlation 
function $\xi(r)$.

\section{Analysis of Data}

In this work we use the most recent spectroscopic 
SDSS data release, DR6 (\cite{dr6}).
We use the same samples and methodology here as presented in \cite{paper1}
of this series. 
LRG's are targeted in the photometric catalog, via cuts in the (g-r, r-i, r) 
color-color-magnitude cube.  We need to
k-correct the magnitudes in order to obtain the absolute magnitudes and
eliminate the brightest and dimmest galaxies. We have seen that the previous
cuts limit the intrinsic luminosity to a range $-23.2<M_r<-21.2$, and we only
eliminate from the catalog some few galaxies that lay out of the limits. Once we
have eliminated these extreme galaxies, we still do not have a volume limited
sample at high redshift. For the 2-point function analysis we  account for 
this using a random catalog with identical selection function but 20 times
denser (to avoid shot-noise) . The same is done in simulations.

There are about $75,000$ LRG galaxies with spectroscopic redshifts 
in the range $z=0.15-0.47$ over $13\%$ of the sky. We break the full
sample into 3 independent subsamples with similar number of galaxies:
low $z=0.15-0.30$, middle $z=0.30-0.40$ and high $z=0.40-0.47$.
In this analysis we will just show results for the  $z=0.15-0.30$ sample
(see \cite{paper1,paper4} for other samples).

To estimate the correlation $\xisp$, 
we use the estimator of \cite{landyszalay},

\begin{equation} \xisp  = \frac{DD - 2DR + RR}{RR} \end{equation}

with a random catalog $N_R=20$ times denser than the SDSS catalog. 
The random catalog has the same redshift (radial) distribution as the data, 
but smoothed with a bin $dz=0.01$ to avoid the elimination of
intrinsic correlations in the data. The random catalog also has the same mask. We count the pairs
in bins of separation along the line-of-sight (LOS), $\pi$,  and across the sky,
$\sigma$. The LOS distance $\pi$ is just the difference between the radial comoving distances
in the pair. The transverse distance $\sigma$ is given by
$\sqrt{s^2-\pi^2}$, where $s$ is the net distance between the pair. We use the small-angle
approximation, as if we had the catalog at an infinite distance, which is accurate
until the angle that separates the galaxy pair in the sky 
is larger than about 10 degrees for 
$\xisp$ (see \cite{szapudiwide} and \cite{matsuwide}). This
condition corresponds to transverse scales larger than $\sigma=80Mpc/h$ for our 
mean catalog.

\begin{figure*}[t]
\includegraphics[scale=.2]{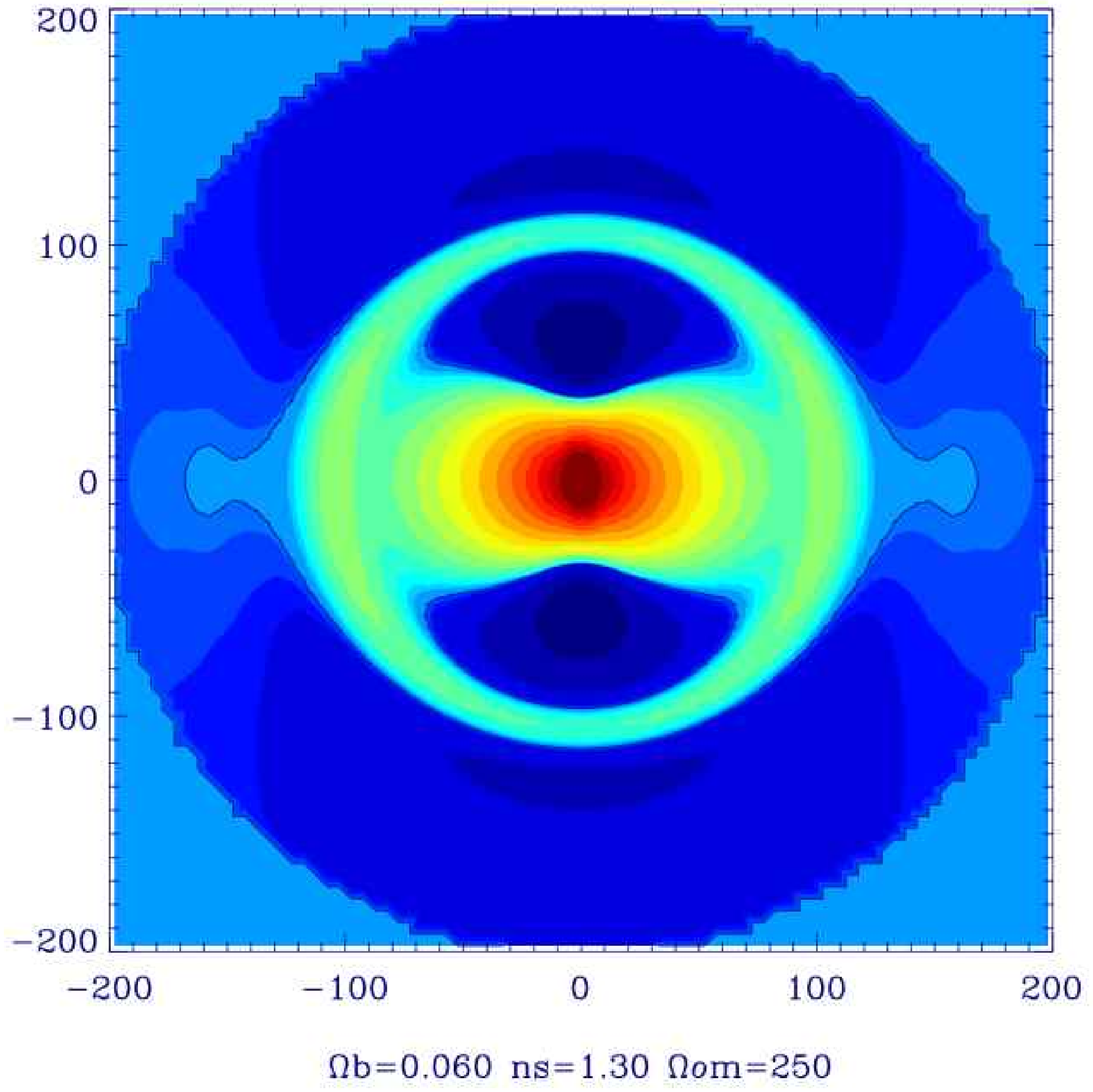}
\includegraphics[scale=.2]{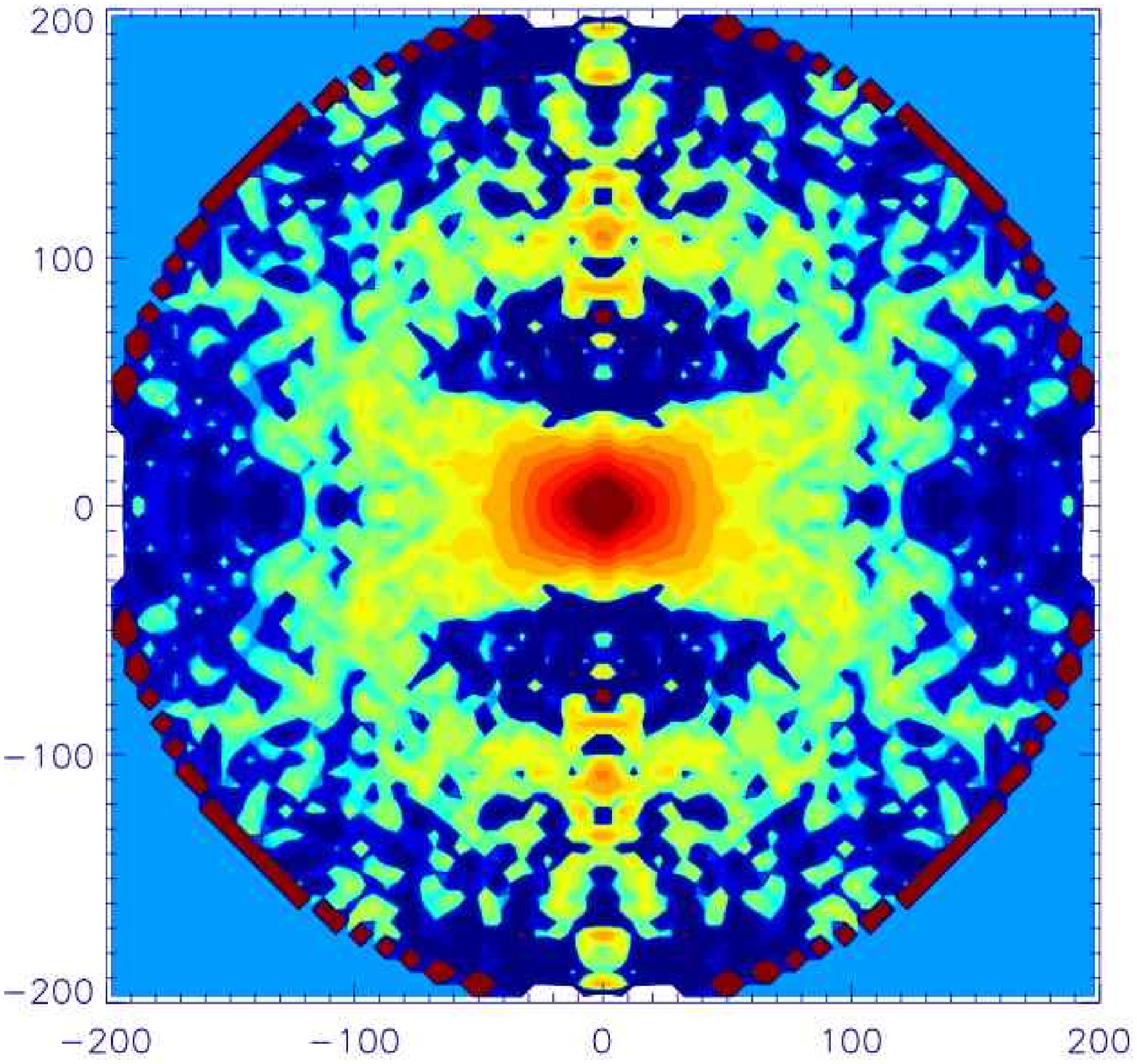}
\includegraphics[scale=.6]{colorbar6.ps}
\caption{A comparison of $\xisp$ in data and models for the z=0.15-0.30 slice.
The vertical axis shows the radial direction, $\pi$, while the horizontal panel shows
the transverse direction $\sigma$.
The left panel includes a model of redshift space distortions which gives
the best fit to the monopole in the galaxy data (see \cite{paper1}). 
Right panel shows the LRG SDSS measurements using the same color scheme. The
data shows a prominent BAO ring at a radius of about $110 \Mpc$, in good agreement with
the model.
\label{fig:pisigma}}
\end{figure*}

The right panel in Fig.\ref{fig:pisigma} shows the measurements
of $\xips$ in the $z=0.15-0.30$ sample. As we will show below there 
is a remarkable agreement with the predictions (left panel)
and there is good evidence for a BAO ring.

\subsection{Errors and simulations}

There are two sources of error or variance in the estimation of
the two-point correlation: a) {\bf shot-noise}
which scales as one over the square root of the number of
pairs in each separation bin b) {\bf sampling variance} which
scales with the amplitude of the correlation.
It is easy to check that for the size and number density of our
sample, the shot-noise term dominates over the sampling variance
error. This has been confirmed in detail by 
using numerical simulations (see \cite{paper1} for more details).
The  simulation contains $2048^3$ dark matter particles, 
in a cube of side $7680\Mpc$  (which we call MICE7680), $\Omega_M=0.25$, 
$\Omega_b =0.044$, $\sigma_8=0.8$, $n_s=0.95$ and $h=0.7$. 
 We have divided this
big cube in $3^3$ cubes of side 2 x 1275Mpc/h, and taking the center of these
secondary cubes as the observation point (as if we were at z=0), we apply the
selection function of LRG, which arrives to z=0.47 (r=1275Mpc/h). We can
obtain 8 octants from the secondary sphere included in the cube, so at the end
we have 8 mock LRG catalogs from each secondary cube, which have the same
density per pixel as LRG in order to have the same level of shot noise, and the
area is slightly smaller (LRG occupies 1/7 of the sky with a different shape).
The final number of independent mock catalogs is $M=216$ (27x8).
 We also apply redshift distortions in the line-of-sight
direction $ s=r+v_{r}/H(z)/a(z)$, using the peculiar velocities $v_r$ from
the simulations. The  error covariance is found from the dispersion of $M$ 
realizations:

\begin{equation}\label{eq:covMC} 
C_{ij} =  \frac{1}{M}
\sum_{k=1}^{M}(\xi(i)^{k}-\widehat\xi(i))(\xi(j)^{k}-\widehat\xi(j))
\end{equation}
where $\xi(i)^{k}$ is the measure in the k-th simulation (k=1,...M) and
$\widehat\xi(i)$ is the mean over M realizations (which we have checked that
agrees with overall mean, indicating that volume effects are small).
The case i=j gives the
diagonal error (variance).  In our analysis  we model the errors using dark matter groups.
These groups are chosen to have the same number density and amplitude of clustering as 
the observed LRG's.  The resulting errorbars from simulations are 
typically in good agreement with Jack-knife errors 
from the actual data (see \cite{paper1} for details).
We have also checked in \cite{paper1} that we can recover the theory 
predictions for $\xisp$ within the errors  by using mock simulations with similar
size as the real data. 

\subsection{Redshift space distortions}

Radial  displacements caused by peculiar velocities
lead to redshift distortions, with two important contributions.
The first, on large scale fluctuations, caused by coherent bulk motion. We see
walls denser and voids bigger and emptier, with a squashing effect in the
2-point correlation function along the line-of-sight: known as 
the Kaiser \cite{Kaiser} effect. At small scales, random velocities 
inside clusters and groups  of galaxies produce a radial
stretching pointed at the observer, known as fingers of God (FOG).
Although such distortions complicate the interpretation of redshift maps as
positional maps, they have the advantage of bearing unique information about the
dynamics of galaxies. In particular, the amplitude of distortions on large
scales yields a measure of the linear redshift distortion parameter $f$.

In the large-scale linear regime, and in the
plane-parallel approximation, the distortion caused by coherent infall velocities takes
a particularly simple form. On average, large scale fluctuations in redshift space $\delta_s$
are enhanced with respect to real space $\delta$ because of the radial velocity infall, so that
$\delta_s \simeq \delta-\theta/3=(1+f/3)\delta$ so that they  are larger by a factor $(1+f/3)$. This enhancement
is anisotropic. In Fourier space:
\begin{equation}\label{eq:kaiserpoint} 
P_s(\vec{k}) = (1 + f \mu_k^2)^2 P(k).
\end{equation} 
where $P(k)$ is the power spectrum of density fluctuations
$\delta$, $\mu$ is the cosine of the angle between $\vec{k}$ and the line-of-sight,
the subscript $s$ indicates redshift space, and $f$ is the 
velocity growth rate in linear theory.

The correlation $\xisp$ is related to the power
spectrum by a Fourier transform:

\begin{equation}
 \xisp=\int P_s(\vec{k})e^{-i\vec{k}\vec{r}}\frac{d^3k}{(2\pi)^3}
\label{eq:fourier}
\end{equation}

 After integration in Eq.[\ref{eq:fourier}], 
these linear distortions in $P_s (\vec{k})$ produce a distinctively anisotropic $\xisp$.
At scales smaller than about $50 \Mpc$ there is a clear squashing in the correlation
function caused by the peculiar velocity divergence $\theta$ field, this effect can be used to estimate $f$, for example
by fitting the normalized quadrupole to the data (eg see \cite{paper1}). For the sample with
z=0.15-0.30  we find $f=0.48-0.83$, which corresponds to
$\Omega_m=0.24-0.32$ when we assume standard gravity ($\gamma=0.55$). 
 
Redshift distortions in the linear regime  produce a lower amplitude and sharper
 baryon acoustic peak in the LOS 
than in the perpendicular direction because 
of the coherent infall into large scale overdensities.
This is illustrated in the left panel of Fig.\ref{fig:pisigma}. A characteristic
feature of this effect is a valley of negative correlations (in blue)
on scales between $\pi = 50-90$ Mpc/h, which is in
very good agreement with our measurements from real SDSS data.
Such a valley is absent without redshift distortions.

\begin{figure}[t]
\sidecaption[t]
\includegraphics[scale=.15]{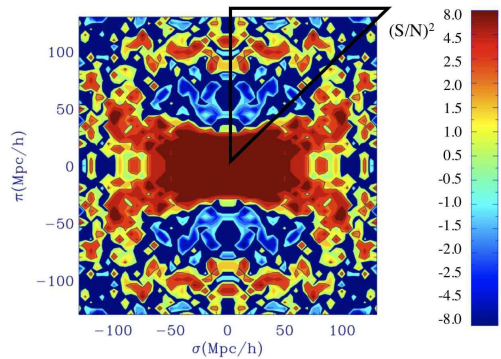}
\caption{Signal-to-noise ratio in $\xisp$
for the z=0.15-0.30 slice. The color scheme denotes
$(S/N)^2$ multiplied by the sign of the signal i.e.
negative values correspond to a negative signal. The triangle
highlights the region $\pi>\sigma$, which receives little weight in the monopole.
The mean $(S/N)^2$ in radial bins in this region is shown in Fig.\ref{fig:s2n.radial}
\label{fig:s2nps.11}}
\end{figure}

In Fig. \ref{fig:s2nps.11}, we show the signal-to-noise of $\xi$ in the
$\sigma-\pi$ plane for the redshift slice $z=0.15 - 0.3$.
This complements the $\xisp$ signal plot in the right panel of Fig. \ref{fig:pisigma}.
The signal-to-noise shown in Fig. \ref{fig:s2nps.11} is for
each pixel of size $5$ Mpc/h by $5$ Mpc/h
(the same pixel size is used in  Fig. \ref{fig:pisigma}). 
Note that there is covariance between pixels, and so 
this figure should be interpreted with some care (see \cite{paper1}).
Nonetheless, it demonstrates the high quality detection of a BAO ring in
the $\sigma-\pi$ plane. The triangle highlights
the region $\pi>\sigma$, which receives not much weight in the monopole, but
where the BAO ring still shows up nicely.
Note that the $(S/N)^2$ shown is modulated by the sign of the signal:
the (blue) valley of negative correlations at $\pi \sim 50 - 90$ Mpc/h
- in accord with the predictions of the Kaiser effect - are detected
with significance as well. The overall coherent structure of a negative valley
before a positive BAO peak (at just the right expected scales) 
is quite striking, and cannot be easily explained away
by noise or systematic effects.

\begin{figure}[t]
\sidecaption[t]
\includegraphics[scale=.15]{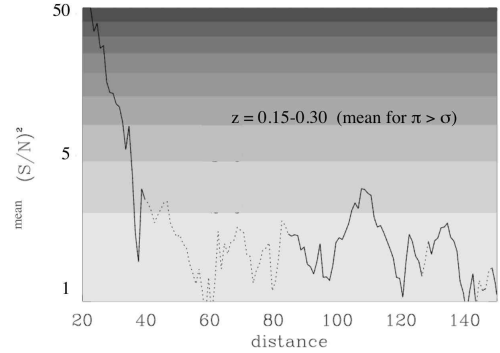}
\caption{Mean square signal-to-noise ratio averaged in radial bins with
$\pi>\sigma$, shown as a function of the radial distance $\pi$. Note that
we include the sign of the signal in doing the mean, see Eq.[\ref{eq:s2nradial}],
which cancels the noise contribution (mean negative values are shown
as a dotted line). 
\label{fig:s2n.radial}}
\end{figure}

The evidence for a BAO peak in the monopole is quite convincing
(see \cite{detection,paper1,paper4}).
The data follows the model prediction and produces a clear
$\Omega_b$ detection which otherwise (without the BAO peak) is
degenerate with other cosmological parameters. But the monopole
signal is dominated by pairs in the perpendicular direction $\sigma>\pi$
and here we would like to assess if the BAO peak is also significant
in the radial direction. We do this by studying the signal-to-noise
ratio in $\xisp$ for $\pi> \sigma$. In
Fig.\ref{fig:s2nps.11} this corresponds to the region inside
the over-plotted triangle. We do the mean signal-to-noise inside the 
region $\pi>\sigma$ as a function of the radius
$s^2=\pi^2+\sigma^2$, in radial shells $d\pm ds$ of width $ds=2.5$ Mpc/h:

\begin{equation}
mean (S/N)^2 = \sum_{s\pm ds} ~(Sign)~  (S/N[\pi,\sigma])^2
\label{eq:s2nradial}
\end{equation}
where $(Sign)$ is the sign of the signal. When the signal is negative
this gives a negative contribution to the mean signal-to-noise square.
If the signal is dominated by noise, positive and negative fluctuations
will tend to cancel and reduced the mean $ (S/N)^2$.
Results are shown in Fig.\ref{fig:s2n.radial}. The mean signal-to-noise
is always larger than unity both in the negative valley between 50-90 Mpc/h
and also around the BAO peak, where the mean $ (S/N)^2$ approaches 2.
This clearly indicates that the BAO peak is also significant in the
radial direction and it also has the shape that is predicted by the
models, with a negative valley  and a positive peak that extend
in a coherent way over the expected lengths. It is unlikely that
noise or systematic errors could reproduce these correlations.
Similar results are found for the other redshift slices, with less
significant detection for the middle slice.

\begin{acknowledgement}
We acknowledge the use of MICE simulations 
(www.ice.cat/mice) developed at the MareNostrum supercomputer
(www.bsc.es) and with support from PIC (www.pic.es),
 the Spanish Ministerio de Ciencia
y Tecnologia (MEC), project AYA2006-06341 with
EC-FEDER funding, Consolider-Ingenio CSD2007-00060
and research project 2005SGR00728
from Generalitat de Catalunya. AC acknowledges support
from the DURSI department of the Generalitat de
Catalunya and the European Social Fund.
\end{acknowledgement}


\end{document}